\documentclass[11pt,twoside]{article}

\usepackage{asp2006}
\usepackage{epsf}
\usepackage{psfig}
\usepackage{lscape}

\begin{document}
\title{Extended Light in E/S0 Galaxies and Implications for Disk Rebirth}   %%% Fill in title
\author{Amanda J. Moffett \altaffilmark{1}, Sheila J. Kannappan \altaffilmark{1}, Seppo Laine \altaffilmark{2}, Lisa H. Wei \altaffilmark{3}, Andrew J. Baker \altaffilmark{4}, Chris D. Impey \altaffilmark{5} }   %%% Fill in author names
%\affil{1: University of North Carolina, 2: Spitzer Science Center, 3: University of Maryland}    %%% Fill in author affiliations

\altaffiltext{1}{University of North Carolina, Chapel Hill, NC 27599, amoffett@physics.unc.edu}
\altaffiltext{2}{Spitzer Science Center, California Institute of Technology, Pasadena, CA 91125}
\altaffiltext{3}{University of Maryland, College Park, MD 20742}
\altaffiltext{4}{Rutgers, the State University of New Jersey, Piscataway, NJ 08854}
\altaffiltext{5}{University of Arizona, Tucson, AZ, 85721}

\begin{abstract} %%% Abstract to run on from here.
The recent discovery of extended ultraviolet (XUV) disks around a large fraction of late-type galaxies provides evidence for unexpectedly large-scale disk building at recent epochs. Combining \emph{GALEX} UV observations with deep optical and \emph{Spitzer} IR imaging, we search for XUV disks in a sample of nearby low-to-intermediate mass E/S0 galaxies to explore evidence for disk rebuilding after mergers. Preliminary visual classification yields ten XUV-disk candidates from the full sample of 30, intriguingly similar to the $\sim$30\% frequency for late-type galaxies. These XUV candidates occur at a wide range of masses
 and on both the red and blue sequences in color vs. stellar mass, indicating a possible association with processes like gas accretion and/or galaxy interactions that would affect the galaxy population broadly.
We go on to apply the quantitative Type 1 and Type 2 XUV-disk definitions to a nine-galaxy subsample analyzed in detail. For this subsample, six of the nine are Type 1 XUVs, i.e., galaxies with UV structure beyond the expected star formation threshold. The other three come close to satisfying the Type 2 definition, but that definition proves problematic to apply to this sample: the NUV-derived star formation threshold radii for our E/S0s often lie inside the 80\% $K_s$-band light ($K_{80}$) radii, violating an implicit assumption of the Type 2 definition, or lie outside but not as far as the definition requires.  Nonetheless, the three otherwise Type 2-like galaxies (``modified Type 2 XUVs'') have higher star formation rates and bluer FUV $-$ NUV colors than the Type 1 XUVs in the sample. We propose that Type 1 XUVs may reflect early or inefficient stages of star formation, while modified Type 2 XUVs perhaps reflect inside-out disk regrowth.
\end{abstract}

%%% MAIN BODY OF TEXT GOES HERE. CONSULT "INSTRUCTIONS FOR AUTHORS USING
%%% LATEX2E MARKUP", SECTIONS 2.3-2.6 FOR HELP WITH EQUATIONS, FIGURES,
%%% AND TABLES.

\section{Background}   %%% Top level section head (remove "%" symbol)

The ability of disk galaxies to survive in a violent universe filled with merger events is one of the central mysteries remaining in our understanding of galaxy evolution. A partial solution to this puzzle may be that while
disks can be destroyed by galaxy-galaxy mergers (yielding E/S0
morphologies), they can also be reborn at a later time. Recent work has identified
a population of low-to-intermediate mass, morphologically defined E/S0 galaxies that are surprisingly blue \citep{KGB},
 indicating young stars. They occupy low density environments and have
substantial cold gas to fuel star formation. Could these galaxies indeed be
regenerating late-type morphology? The recent \emph{GALEX} discovery of
extended ultraviolet (XUV) disks (e.g., \citealp{Th05}; \citealp{Gil05}),
representing very young stars around late-type galaxies, has
provided tantalizing evidence for large-scale disk building at $z=0$. Here we search for
XUV emission in E/S0 galaxies to examine evidence for disk regrowth after mergers.
Our sample includes 30 E/S0s selected to be representative of the red and blue sequences in a stellar
mass range primarily below  $\sim$5 x 10$^{10}$~M$_{\odot}$ (Fig. 1), where many E/S0s have substantial gas. Nine of these have
fully reduced data, enabling detailed analysis.

\begin{figure}[!t]
\epsscale{0.7}
\plotone{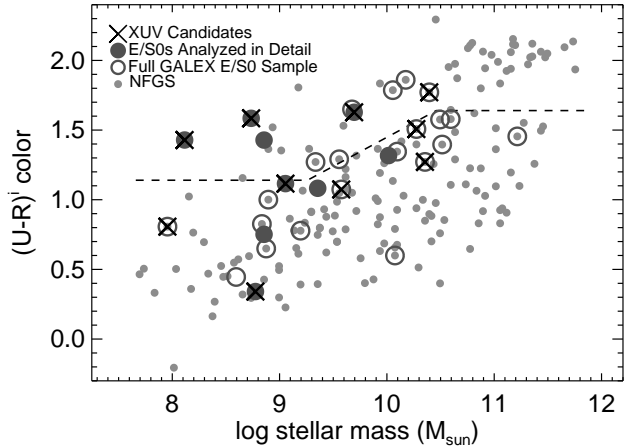}
\caption{Our E/S0 sample in color-stellar mass space. The dashed line divides the red and blue sequences. The full 30-galaxy E/S0 sample is denoted by large filled and open circles, the ten visually identified XUV-disk candidates in this sample by crosses, and the nine-galaxy subsample considered in detail by large filled circles. The small grey dots indicate objects in the Nearby Field Galaxy Survey \citep{Jan}, the source of most of our E/S0s. }
\end{figure}

\section{Type 1 and 2 XUV disks}

\citet{Th07} define XUV disks of two types. A Type 1 XUV disk has more than one structured UV-bright emission complex beyond a central surface brightness contour corresponding to the anticipated star formation threshold (NUV 27.35 AB mag\,arcsec$^{-2}$). A Type 2 XUV disk has FUV(AB) $-$ $K_s$(AB) $\leq$ 4 in a ``large'' (i.e., area at least 7 times the enclosed area of the $K_{80}$ contour) optically low surface brightness (LSB) zone within the expected star formation threshold but outside $K_{80}$. At present we can apply Thilker's definitions to only the nine galaxies with fully analyzed data, so we apply a third, more subjective identification method to our full 30-galaxy E/S0 sample in \S 3.

Applying the Type 1 XUV-disk definition to our subsample yields a 6/9 or
$\sim$67\% incidence of XUV disks of this type and one borderline case with fainter extended emission. 
An example of a Type 1 XUV disk is NGC 4117 (Fig. 2 left panels), one of several identified on the red sequence. Its NUV imaging shows 
clumpy UV morphology outside the anticipated star formation threshold. We note, however, that the Type 1 requirement of structured emission is sometimes difficult to apply consistently to galaxies like those in our sample, which have a smaller extent on the sky
than those considered in Thilker's defining sample, so their structure is blurred by the relatively low angular resolution of \emph{GALEX}.

\begin{figure}[t!]
\epsscale{1}
\plotone{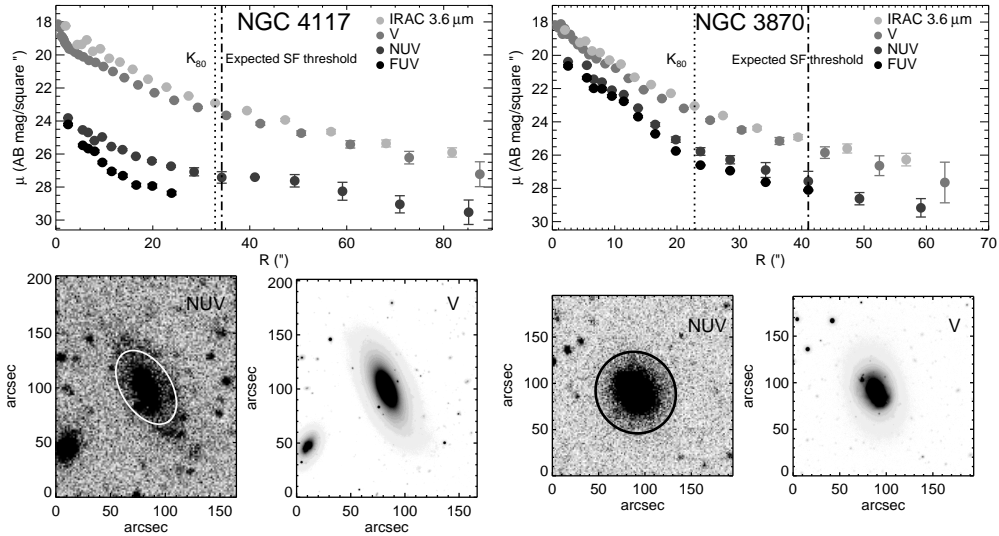}
\caption{\emph{Left Panels} - NGC 4117 - Top: Multi-wavelength surface brightness profiles. Bottom: \emph{GALEX} NUV with overlay of the expected star formation threshold and Mosaic $V$ with matched spatial scale.
\emph{Right Panels} - NGC 3870 - Same as for NGC 4117.}
\end{figure}

The implicit assumption of the Type 2 XUV-disk definition, that the expected star formation threshold lies outside
the $K_{80}$ contour, is satisfied in our sample primarily for E/S0s with ``spiral-like'' or even bluer colors, for example NGC 3870 (Fig. 2 right panels). This definition's lack of applicability to many galaxies in our sample is intriguing. Figure 3 shows that as the ratio between the $K_{80}$ radius and the expected star formation threshold radius increases for our sample objects, indicating centrally concentrated star formation, the star formation rate tends to decrease. Conversely, three of our UV-bluest E/S0s (NGC 3870, NGC 3011, and IC 692, the borderline Type 1 case) do have their star formation thresholds outside $K_{80}$ and satisfy the Type 2 requirement of FUV - $K_s$ $\leq$ 4 in the LSB zone between the $K_{80}$ and threshold radii (converting from IRAC 3.6 $\mu$m to $K_s$-band intensity following \citeauthor{Le08} \citeyear{Le08}). Although the LSB zone is not ``large'' as for a standard Type 2 XUV, these galaxies have bluer UV colors and higher star formation rates than our Type 1 XUVs (Fig. 3), so we label them ``modified Type 2'' XUVs. Within the constraints of small number statistics, these results suggest that Type 1 morphology may be indicative of inefficient star formation or star formation just beginning, while the bluer disks forming in modified Type 2 XUVs may reflect inside-out disk regrowth.

\newpage

\section{Continuing Work: The Full Sample}

\begin{figure}[t!]
\epsscale{1.}
\plotone{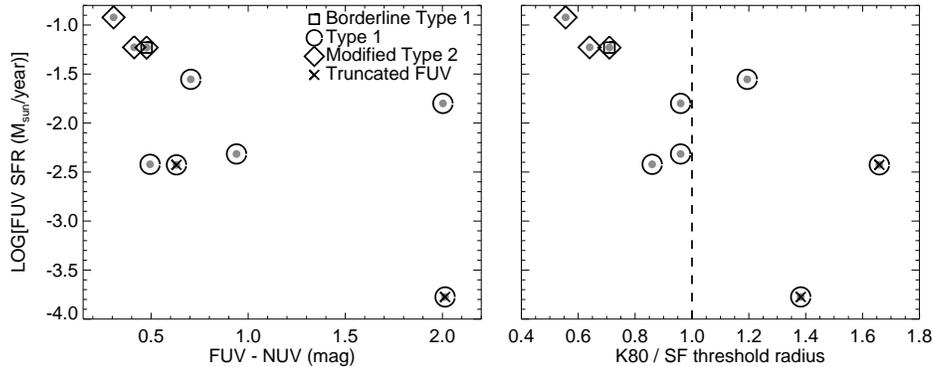}
\caption{\emph{Left} - FUV-derived star formation rate versus FUV $-$ NUV color, both measured within the radius of the last detected point in the FUV profile. \emph{Right} - SFR vs.~the ratio of the $K_{80}$ radius to the expected star formation threshold radius. The dashed line marks a ratio of one. Only foreground extinction corrections have been applied. }
\end{figure}

As a precursor to completing the data analysis necessary to apply Thilker's Type 1 and 2 definitions to our full 30-galaxy sample, we apply a third, more subjective identification method.
An XUV disk in this third method is defined as having extended UV emission relative to the optical in a visual comparison of NUV and SDSS $g$-band images. 
By this preliminary identification $\sim$33\% of our sample are XUV candidates, similar to the $\sim$30\% fraction for late-type galaxies (\citealp{Th07}). For the nine-galaxy
subsample analyzed in detail, this method yields 5/9 or
$\sim$56\% incidence of XUV disks, which is actually lower than the frequency of XUVs identified using Thilker's Type 1 and 2 criteria. Our preliminary identification of XUV-disk morphologies in E/S0s at a wide range of masses
 and on both the red and blue sequences may indicate an association with processes like gas accretion and/or galaxy interactions that affect the galaxy population broadly, perhaps contributing to disk regrowth after mergers.

%\subsection{}   %%% Second level section head (remove "%" symbol)
%\subsubsection{}   %%% Lowest level section head (remove "%" symbol)
%\section*{}    %%% Unnumbered top level section head (remove "%" symbol)
%\subsection*{}   %%% Unnumbered second level section head (remove "%" symbol)

\acknowledgements %%% Text of acknowledgements runs on after this command.
We thank Stuart Vogel (University of Maryland), Jocelly Guie (University of Texas at Austin), Hillary Mathis (NOAO), and Shardha Jogee (University of Texas at Austin) for their role in acquiring the data. We acknowledge support from \emph{GALEX} grant NNX07AT33G.

%%% THE BIBLIOGRAPHY
%%%
%%% CONSULT SECTION 3 OF "INSTRUCTIONS FOR AUTHORS" FOR HOW TO USE NATBIB.
%%% AUTHORS ARE ENCOURAGED TO USE EITHER THE "THEBIBLIOGRAPY" ENVIRONMENT
%%% BY UNCOMMENTING (DELETING THE "%" SYMBOL) THE COMMANDS BELOW, OR BY
%%% USING THE BIBTEX ENVIRONMENT. TO FIND OUT WHICH IS APPLICABLE TO YOUR
%%% CONTRIBUTION, CONSULT THE VOLUME EDITORS FOR YOUR PROCEEDINGS.
%%%

\end{document}